\documentstyle[12pt,psfig]{article}
\makeatletter
\def\@cite#1#2{{[{#1}]\if@tempswa\typeout
{IJCGA warning: optional citation argument
ignored: `#2'} \fi}}

\newcount\@tempcntc
\def\@citex[#1]#2{\if@filesw\immediate\write\@auxout{\string\citation{#2}}\fi
  \@tempcnta\z@\@tempcntb\m@ne\def\@citea{}\@cite{\@for\@citeb:=#2\do
    {\@ifundefined
       {b@\@citeb}{\@citeo\@tempcntb\m@ne\@citea\def\@citea{,}{\bf ?}\@warning
       {Citation `\@citeb' on page \thepage \space undefined}}%
    {\setbox\z@\hbox{\global\@tempcntc0\csname b@\@citeb\endcsname\relax}%
     \ifnum\@tempcntc=\z@ \@citeo\@tempcntb\m@ne
       \@citea\def\@citea{,}\hbox{\csname b@\@citeb\endcsname}%
     \else
      \advance\@tempcntb\@ne
      \ifnum\@tempcntb=\@tempcntc
      \else\advance\@tempcntb\m@ne\@citeo
      \@tempcnta\@tempcntc\@tempcntb\@tempcntc\fi\fi}}\@citeo}{#1}}
\def\@citeo{\ifnum\@tempcnta>\@tempcntb\else\@citea\def\@citea{,}%
  \ifnum\@tempcnta=\@tempcntb\the\@tempcnta\else
   {\advance\@tempcnta\@ne\ifnum\@tempcnta=\@tempcntb \else
\def\@citea{--}\fi
    \advance\@tempcnta\m@ne\the\@tempcnta\@citea\the\@tempcntb}\fi\fi}
\makeatother
\newenvironment{Eqnarray}%
     {\arraycolsep 0.14em\begin{eqnarray}}{\end{eqnarray}}

\def\simlt{\stackrel{<}{{}_\sim}}
\def\simgt{\stackrel{>}{{}_\sim}}
\def\be{\begin{equation}}
\def\ee{\end{equation}}
\def\bear{\be\begin{array}}
\def\eear{\end{array}\ee}
\def\bea{\begin{Eqnarray}}
\def\eea{\end{Eqnarray}}

\def\lsim{\mathrel{\raise.3ex\hbox{$<$\kern-.75em\lower1ex\hbox{$\sim$}}}}
\def\gsim{\mathrel{\raise.3ex\hbox{$>$\kern-.75em\lower1ex\hbox{$\sim$}}}}
\def\ifmath#1{\relax\ifmmode #1\else $#1$\fi}
\def\ls#1{\ifmath{_{\lower1.5pt\hbox{$\scriptstyle #1$}}}}

\def\beq{\begin{equation}}
\def\eeq{\end{equation}}
\def\beqa{\begin{Eqnarray}}
\def\eeqa{\end{Eqnarray}}

\def\baselinestretch{1}
\begin{document}
\def\IJMPA #1 #2 #3 {{\sl Int.~J.~Mod.~Phys.}~{\bf A#1}\ (19#2) #3$\,$}
\def\MPLA #1 #2 #3 {{\sl Mod.~Phys.~Lett.}~{\bf A#1}\ (19#2) #3$\,$}
\def\NPB #1 #2 #3 {{\sl Nucl.~Phys.}~{\bf B#1}\ (19#2) #3$\,$}
\def\PLB #1 #2 #3 {{\sl Phys.~Lett.}~{\bf B#1}\ (19#2) #3$\,$}
\def\PR #1 #2 #3 {{\sl Phys.~Rep.}~{\bf#1}\ (19#2) #3$\,$}
\def\JHEP #1 #2 #3 {{\sl JHEP}~{\bf #1}~(19#2)~#3$\,$}
\def\PRD #1 #2 #3 {{\sl Phys.~Rev.}~{\bf D#1}\ (19#2) #3$\,$}
\def\PTP #1 #2 #3 {{\sl Prog.~Theor.~Phys.}~{\bf #1}\ (19#2) #3$\,$}
\def\PRL #1 #2 #3 {{\sl Phys.~Rev.~Lett.}~{\bf#1}\ (19#2) #3$\,$}
\def\RMP #1 #2 #3 {{\sl Rev.~Mod.~Phys.}~{\bf#1}\ (19#2) #3$\,$}
\def\ZPC #1 #2 #3 {{\sl Z.~Phys.}~{\bf C#1}\ (19#2) #3$\,$}
\def\PPNP#1 #2 #3 {{\sl Prog. Part. Nucl. Phys. }{\bf #1} (#2) #3$\,$}

\catcode`@=11
\newtoks\@stequation
\def\subequations{\refstepcounter{equation}%
\edef\@savedequation{\the\c@equation}%
  \@stequation=\expandafter{\theequation}
  \edef\@savedtheequation{\the\@stequation}
  \edef\oldtheequation{\theequation}%
  \setcounter{equation}{0}%
  \def\theequation{\oldtheequation\alph{equation}}}
\def\endsubequations{\setcounter{equation}{\@savedequation}%
  \@stequation=\expandafter{\@savedtheequation}%
  \edef\theequation{\the\@stequation}\global\@ignoretrue

\noindent}
\catcode`@=12

\begin{titlepage}

\title{{\bf Impact of radiative corrections\\on sterile neutrino scenarios}}
\vskip2in
\author{ 
{\bf A. Ibarra$$\footnote{\baselineskip=16pt  E-mail: {\tt
alejandro@makoki.iem.csic.es}}}  and 
{\bf I. Navarro$$\footnote{\baselineskip=16pt E-mail: {\tt
ignacio@makoki.iem.csic.es}}}\\ 
\hspace{3cm}\\
 ~{\small Instituto de Estructura de la Materia (CSIC)}\\
{\small  Serrano 123, 28006 Madrid, Spain}
\hspace{0.3cm}\\
} 
\date{}
\maketitle 
\def\baselinestretch{1.15} 
\begin{abstract}
\noindent
In sterile neutrino scenarios, radiative corrections induce mass splittings proportional to the top Yukawa coupling, in contrast to the three active neutrino case where the induced splittings are proportional to the tau Yukawa coupling. In view of this, we have analyzed the stability of the four-neutrino schemes favored by oscillation experiments, consisting in two pairs of nearly degenerate neutrinos separated by the LSND gap. Requiring compatibility with the measurements of the abundances of primordial elements produced in Big Bang Nucleosynthesis, we find that when the heaviest pair corresponds to the solar neutrinos (mainly an admixture of $\nu_e -\nu_s$) the natural mass splitting is 3-5 orders of magnitude larger than the observed one, discrediting the scenario from a theoretical point of view. On the contrary, the scheme where the heaviest pair corresponds to the atmospheric neutrinos (mainly an admixture of $\nu_{\mu} -\nu_{\tau}$) is safe from radiative corrections due to the small sterile component of these mass eigenstates.
\end{abstract}

\thispagestyle{empty}
\leftline{IEM-FT-199/99}
\leftline{December 1999}
\leftline{}


\end{titlepage}
\setcounter{footnote}{0} \setcounter{page}{1}
\newpage
\baselineskip=20pt

\noindent

\section{Introduction}

Strong indications of neutrino oscillations have been gathered in the atmospheric and solar neutrino experiments \cite{SK,ATM,SOL}. These data point to two distinct mass squared differences: they range from $5\times 10^{-4}-10^{-2}$ $ {\rm eV^2}$ for the atmospheric oscillation to $5\times 10^{-11}-2\times 10^{-4}$ ${\rm eV^2}$ for the solar one. There is further indication of neutrino oscillation in the, so far unconfirmed, LSND data \cite{LSND}, which would require a very distinct mass difference: $0.3-1$ ${\rm eV^2}$. In order to accommodate all mass squared differences, at least four light species of neutrinos are needed, one of which must be sterile, i.e., singlet under the Standard Model (SM) gauge group \cite{LEP,sterile}. 

Among all neutrino spectra involving one sterile neutrino there are two somewhat preferred by the combined analysis of solar, atmospheric and LSND experiments \cite{bilenky}. These schemes are characterized by two pairs of nearly degenerate neutrinos, accounting for the solar and atmospheric oscillations, separated by a gap of $O(1 \hspace{0.1cm}{\rm eV})$, responsible of the LSND anomaly (see fig.(1)). In one of these (scheme A), the two heaviest neutrinos correspond basically to the solar neutrinos, and in the other (scheme B), these are mainly the atmospheric neutrinos. The atmospheric signal is interpreted as a $\nu _{\mu} - \nu _{\tau}$ oscillation while the solar deficit is explained as a $\nu _{e} -\nu _{s}$ oscillation. In this paper, we will restrict ourselves to these scenarios. It is important to point out that when the solar neutrino problem is interpreted as $\nu _{e}-\nu _{s}$ oscillations, only the small angle MSW solution (SAMSW) is allowed \cite{NLA}, yielding the fits $3\times 10^{-6}\hspace{0.1cm} {\rm eV^2}< \Delta m_{sol} ^2 <10^{-5} \hspace{0.1cm} {\rm eV^2}$ and  $2\times 10^{-3} < \sin^2 2\theta_{sol} < 2\times 10^{-2}$. 

Here we will consider a scenario with a $4\times 4$ mass matrix for the three active neutrinos plus the sterile one. We define the effective mass term for the four species of neutrinos in the flavor basis as 
\bea 
{\cal L}=-\frac{1}{2} \nu^T {\cal M_\nu}  \nu\;+\;{\rm h.c.}
\label{Mnu}
\eea
where $\nu^T =(\nu_e, \nu_{\mu}, \nu_{\tau}, \nu_{s})$ and ${\cal M_\nu}$ is the neutrino mass matrix. This is diagonalized in the usual way 
\bea 
{\cal M_\nu} = V^* D V^\dagger
\eea
where V is a unitary $4\times 4$ matrix relating flavor eigenstates to mass eigenstates 
\bea \pmatrix{\nu_e \cr \nu_\mu\cr
\nu_\tau\cr \nu_s \cr} = V \pmatrix{\nu_1\cr \nu_2\cr \nu_3\cr \nu_4} 
\eea
and $D=diag(m_1 e^{i \phi _1}, m_2 e^{i \phi _2}, m_3 e^{i \phi _3}, m_4)$, with all $m_i \geq 0$. It will be useful in the analysis of the complex case to absorb the Majorana phases in the Maki-Nakagawa-Sakata (MNS) matrix $U$ \cite{MNS}, defined as
\bea
U=V\cdot diag(e^{-i \phi _1/2},e^{-i \phi _2/2},e^{-i \phi _3/2},1)
\eea
so that $U^T {\cal M_\nu} U = diag(m_1, m_2, m_3, m_4)$.

\begin{figure}
\centerline{\vbox{
\psfig{figure=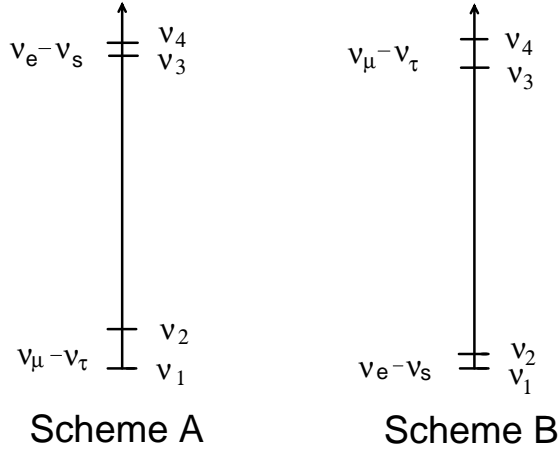,height=8.5cm,width=8.5cm,bbllx=3.cm,%
bblly=1.cm,bburx=18.cm,bbury=16.cm}}}
\caption
{\footnotesize
Schemes for the neutrino spectrum}
\end{figure}

We will consider just the SM fields plus one singlet fermion, to be regarded as the sterile neutrino (the extension to the Minimal Supersymmetric Standard Model (MSSM)  is straightforward). The Majorana mass matrix for the three active neutrinos and the sterile one is formed by three kinds of terms. The active-active entries of the mass matrix come from a dimension five operator \cite{weinberg}
\be
\delta{\cal L}=-\frac{1}{4}\kappa_{ij} (H\cdot L_{i})(H\cdot L_{j}) + {\mathrm h.c.},
\ee
where $H$ is the SM Higgs, $L_i$ ($i=e,\mu ,\tau $) are the lepton doublets and $\kappa_{ij}$ is a $3\times 3$ symmetric matricial coupling. There is also a purely Majorana mass term for the sterile neutrino

\be
\delta{\cal L}=-\frac{1}{2} \nu_{s} m_{s} \nu_{s} + {\mathrm h.c.},
\ee
and Yukawa couplings, responsible for the active-sterile mixing
\be
\delta{\cal L}= y_{i} \nu_{s} H\cdot L_{i} + {\mathrm h.c.}.
\ee
After electroweak symmetry breaking, a $4\times 4$ mass matrix for the four species of neutrinos arises 
\be
{\cal M}_{\nu}= \pmatrix{\kappa v^2 /2 &-{\bf y_{\nu}^{T}} v/\sqrt{2}\cr
-{\bf y_{\nu}} v/\sqrt{2} & m_{s} \cr}
\ee
where ${\bf y_{\nu}}=(y_{e},y_{\mu},y_{\tau})$ are the neutrino Yukawa couplings, and $v=246$ GeV. Note that all the entries of ${\cal M}_{\nu}$ must be at most of $O$(1 eV) to obtain a significant mixing between the sterile and the active neutrinos \footnote{This condition has usually been claimed to be rather unnatural. However, see \cite{langacker} for a framework derived from string theories in which this requirement can be fulfilled naturally.}. The parameters appearing in the Lagrangian that form the neutrino mass matrix have a different renormalization group behavior, as can be checked directly from their renormalization group equations (RGEs) \cite{SMrge,cein1}
\bea 16\pi^2 \frac{d \kappa}{dt}= \left[-3g_2^2+2\lambda+2{\mathrm S} \right]\kappa -\frac{1}{2}\left[\kappa({\bf
Y_e^\dagger Y_e + y_{\nu}^\dagger y_{\nu} }) + ({\bf Y_e^\dagger Y_e + y_{\nu}^\dagger y_{\nu} })^T\kappa\right]\,
\label{rg1}
\eea
\bea
\label{rg2}
\frac{d {\bf y_{\nu}}}{dt}= -\frac{1}{16\pi^2} {\bf y_{\nu} }\left[\left(
\frac{9}{4}g_2^2+\frac{3}{4}g_1^2-{\mathrm S}
\right){\bf I_3}-\frac{3}{2}\left( {\bf
y_{\nu}^\dagger  y_{\nu}}-{\bf Y_e^\dagger Y_e}\right) \right],
\eea 
\bea
\label{rg3}
\frac{d m_{s}}{dt}=\frac{1}{8\pi^2} m_{s} {\bf y_{\nu}}{\bf y_{\nu}^\dagger},
\eea 
where
\bea
{\mathrm S}={\mathrm Tr}(3{\bf Y_U^\dagger Y_U}
+3 {\bf Y_D^\dagger Y_D}+{\bf y_{\nu}^\dagger y_{\nu}}+{\bf Y_e^\dagger Y_e}),
\eea 
and $t=\log \mu$. In the formulas above, $g_2$ and $\lambda$ are the
$SU(2)$ gauge coupling and the quartic Higgs coupling respectively, and ${\bf Y_{U,D,e}}$ are
the Yukawa matrices for up quarks, down quarks and charged leptons. Previous papers \cite{cein1,ellos,ellis,cein2,cein3,BRS,haba,ma,poko,cein4} have considered the modification of the mixing angles and mass splittings due to the last term in eq.(\ref{rg1}) in scenarios with three active neutrinos. The first term in eq.(\ref{rg1}), which depends on $h_t$, produces a flavor independent modification of the active neutrino mass matrix. However, in eq.(\ref{rg2}) the dependence on $h_t$ is different, while eq.(\ref{rg3}) lacks a $h_t$ dependent term, thus important modifications of the mass texture are expected now. This will have important consequences on the theoretical plausibility of some kinds of neutrino spectra. The aim of this paper is to impose naturalness constraints on sterile neutrino scenarios from radiative corrections. This approach has already been employed for the three active neutrinos scenario \cite{cein1,ellos,ellis,cein2,cein3,BRS,haba,ma,poko,cein4}, but now the effects are expected to be larger, as a consequence of the different nature of the sterile neutrino. Furthermore, similar strong renormalization effects are expected in more general models containing sterile neutrinos.

Recalling that the Yukawa coupling between the active and the sterile neutrinos must be small, in order to have masses of $O$(1 eV), in a first approximation we can neglect all couplings but $h_t$ in eqs.(\ref{rg1},\ref{rg2},\ref{rg3}). In this approximation, the RGE for the neutrino mass matrix is
\bea
\label{rg4}
\frac{d {\cal M_{\nu}}}{dt}\simeq \frac{1}{16\pi^2} \left[ {\cal M_{\nu}} \pmatrix{3h_{t}^2 {\bf I_{3}} & 0 \cr
0 & 0 \cr} + \pmatrix{3h_{t}^2 {\bf I_{3}} & 0 \cr
0 & 0 \cr}  {\cal M_{\nu}} \right].
\eea

It is worth noticing that the RGE for the mass matrix has {\it formally} the same structure as the one considered in previous works for the three neutrino case (see eq.(14) in \cite{cein4}) with $P= -{1 \over {16 \pi^2}} \pmatrix{3h_{t}^2 {\bf I_{3}} & 0 \cr
0 & 0 \cr}$ and $\kappa_U=0$. Therefore, we can apply the general results derived there to obtain the RGEs for the mass eigenvalues
\be
\label{RGmass}
\frac{d m_i}{dt} \simeq  \frac {3 h_{t}^2}{8 \pi^2} (1 - |U_{si}|^2) \;m_{i} ,
\ee
and the MNS matrix
\bea 
\frac{d U}{dt} \simeq U T,
\label{URG}
\eea
with $T$ a $4\times 4$ matrix defined as
\bea
\label{Tgendef}
T_{ii}&\equiv&0,\nonumber\vspace{.5cm}\\
T_{ij}&\equiv&\frac {3 h_t^2}{16 \pi^2} \left[\;\frac{m_i+m_j}{m_i-m_j} \;Re(U_{si}^*U_{sj})+i\frac{m_i-m_j}{m_i+m_j} \;Im(U_{si}^*U_{sj})\right],\hspace{1cm} i\neq j,
\eea
with $i=1, 2, 3, 4$. Starting with degenerate neutrino masses at a high energy scale $\Lambda$, the degeneracy will be lifted at $M_Z$ by terms proportional to the top Yukawa coupling, in a way that depends on the last row of the MNS matrix. Strictly speaking, when $m_{i}=m_{j}$, there would be a divergence in eq.(\ref{URG}), but there is also an ambiguity in the definition of the MNS matrix. It can be shown \cite{cein4} that, once the ambiguity is removed by the running, the mixing matrix satisfies $Re(U_{si}^*U_{sj})=0$, making eq.(\ref{URG}) finite.

For the MSSM case, the relevant RGEs for this analysis can be found in \cite{cein2,rgesusy}, although there is a complete analogy with the SM case: neglecting all couplings but the top Yukawa coupling, eq.(\ref{RGmass}) is still valid with the replacement $h_{t} \rightarrow h_{t} /sin\beta$.

In view of the form of the RGEs for the masses and mixing angles, any information about the $U_{si}$ elements of the MNS matrix would be certainly valuable. Fortunately, the remarkable agreement between the predictions of Big Bang Nucleosynthesis (BBN) and the observed abundance of primordial light elements sets stringent bounds on $U_{si}$. The upper bound $N_{\nu}^{BBN}<4$ for the effective number of neutrinos implies that, for scheme A  \cite{BBN}
\bea
|U_{s3}|^2 + |U_{s4}|^2 \simgt 0.99,
\label{BBNA}
\eea
while for scheme B
\bea
|U_{s3}|^2 + |U_{s4}|^2 \simlt 10^{-4}.
\label{BBNB}
\eea

These constraints will have important consequences in the analysis to follow.

\section{The effect of radiative corrections}

As explained in the introduction, radiative corrections play an important role when two mass eigenstates are almost degenerate, i.e., $\Delta m_{ij}^2 \ll m^{2}_{i,j}$. If this is the case, the renormalization group evolution can produce mass differences larger than the observed ones, requiring a fine tuning of the neutrino parameters  at high energy to achieve a correct splitting. This fact was exploited in \cite{cein3,BRS} to impose severe constraints on the vacuum oscillation solution to the solar neutrino problem. In this paper we are going to perform a similar analysis for the experimentally favored neutrino patterns presented in the introduction.

We will assume that at some high energy scale $\Lambda$ new physics generates a neutrino mass texture ${\cal M}_{\nu} (\Lambda)$ in which the heaviest pair of neutrinos have masses nearly degenerated and the other two are $\sim 0$ (we neglect the effects of the RGEs on the masses of the two lightest eigenstates because a hierarchical spectrum will not be essentially modified by the running). After integration from $\Lambda$ to $M_Z$, the radiatively corrected mass eigenvalues can be obtained  from eq.(\ref{RGmass}), giving
\bear{cl}
m_1(M_Z) \sim & 0,\vspace{0.2cm}\\
m_2(M_Z) \sim & 0,\vspace{0.2cm}\\
m_3(M_Z) \simeq & m_3 (\Lambda) ~ I_t ~ I_{s3},\vspace{0.2cm}\\
m_4(M_Z) \simeq & m_4 (\Lambda) ~ I_t ~ I_{s4},
\label{eigensame}
\eear
where 
\bea
I_t\equiv\exp\left(-\int_{t_Z}^{t_\Lambda} \frac{3 h_t^2(t^\prime)}{8 \pi^2} dt^\prime\right), ~~~~
I_{si} \equiv\exp\left(\int_{t_Z}^{t_\Lambda} \frac{3 h_t^2(t^\prime)}{8 \pi^2} |U_{si}|^2(t^\prime)dt^\prime\right).
\label{integ}
\eea
The typical size of $I_t$ is $\sim 0.93$ ($\sim 0.67$) for $\Lambda=10^3$ GeV ($10^{12}$ GeV) in the SM, and  $\sim 0.52$ for $\Lambda=10^{12}$ GeV in the MSSM with $\tan \beta=2$.

In the following subsections, we will analyze two different patterns, labeled as schemes A and B. In scheme A, the heaviest neutrinos participate in solar oscillations, and thus the mass splitting between them is small. In addition, these mass eigenstates have a large sterile component. Therefore, it is in this scheme where larger renormalization effects are expected and will be studied in detail, both the real and the complex case. On the other hand, in scheme B the heaviest pair of neutrinos has a small sterile component and the required splitting between them is larger, so in this case radiative corrections are not expected to have dramatic consequences.

\subsection{Scheme A}

\subsubsection{Real Case}
Throughout the analysis of the real case we will work for convenience with the $V$ matrix, setting all the CP violating phases in $V$ to $0$ and allowing the Majorana phases to be either $0$ or $\pi$, yielding positive or negative mass eigenvalues, respectively. In this case the renormalization effects are sensitive to the relative sign of the mass eigenvalues. In consequence, we will consider the two possibilities separately: $m_{3} \sim m_{4}$ and $m_{3} \sim -m_{4}$.

\vspace{0.5cm}
$a)$ $m_{3} \simeq m_{4}$
\vspace{0.5cm}

The solar splitting produced in scheme A by the renormalization group evolution can be obtained from eqs.(\ref {eigensame}), giving
\bea
\Delta m_{sol}^2\simeq m_0^2 ~ I_t^2 I_{s4}^2 (1- I_{s3}^2 I_{s4}^{-2}),
\label{oscilA1}
\eea
where $m_0^2 I_{t}^2 I_{s4}^2 =m_4^2(M_Z) \simeq \Delta m_{LSND}^2 \sim$ 1 eV to account for the LSND anomaly.

The degeneracy of the mass eigenstates at $\Lambda$ has also an important impact on the radiative corrections of the mixing matrix. The renormalization group equation of the MNS matrix, given by eq.(\ref{URG}), is dominated in this case by 
\bea
T_{43}=\frac{3 h_t^2}{16 \pi^2} \frac{m_{4}+m_{3}}{m_{4}-m_{3}}  V_{s4} V_{s3}.
\label{Tdomin}
\eea
Thus, the renormalization group evolution will quickly drive $V_{s3}\rightarrow 0$  in the infrared \cite{cein4}. Therefore, the BBN constraint (\ref{BBNA}) implies that $V_{s4}^2(M_Z) \simgt 0.99$. This value is close to the fixed point $V_{s4}(M_Z)=1$ and will not change appreciably with the scale, hence, it can be considered constant. In this approximation, and neglecting terms of the order of $1-V_{s4}^2$, eq.(\ref {oscilA1}) reads
\bea
\Delta m_{sol}^2\simeq \Delta m_{LSND}^2 (1-I_t^{2 \,V_{s4}^2} ) \sim \Delta m_{LSND}^2  (1-I_t^2).
\label{oscilA2}
\eea  
For example, taking $\Delta m_{LSND}^2\simeq 0.3$ ${\rm eV^2}$ the induced splitting is $\Delta m_{sol}^2\simeq 0.04$ ${\rm eV^2}$($0.17$ ${\rm eV^2}$) for $\Lambda\simeq 10^3$ GeV ($10^{12}$ GeV) in the SM, and $\Delta m_{sol}^2\simeq 0.22$ ${\rm eV^2}$ for $\Lambda\simeq 10^{12}$ GeV in the MSSM with $\tan \beta=2$, which are too large compared to {\it any} explanation of the solar neutrino deficit by several orders of magnitude. 

The only way out of this shortcoming would be an extreme fine tuning between the initial values of the mass splittings and the effect of the RGEs. To obtain the correct value for $\Delta m_{sol}^2 $ at $M_Z$, the mass splitting at the scale $\Lambda$ is forced to lie in the narrow range
\bea
|m_4^2(\Lambda)-m_3^2(\Lambda)| \sim |(I_t^{-2}-1)\Delta m_{LSND}^2 \pm \Delta m_{sol}^2|,
\label{tuning}
\eea
where we have neglected again terms of the order of $1-V_{s4}^2$. In the most favorable choice of scenario and parameters for the fine-tuning problem, namely the SM with $\Lambda=10^3 \,{\rm GeV}$ and $\Delta m_{LSND}^2= 0.3 \,{\rm eV^2}$, we get from (\ref{tuning}) $|m_4^2(\Lambda)-m_3^2(\Lambda)| \sim (2.309 \pm 0.001) \times 10^{-2}\; {\rm eV^2}$, i.e. the initial splitting has to be adjusted in less than one part in $10^3$. Raising the values of $\Delta m_{LSND}^2$ or $\Lambda$, or going to the supersymmetric case, requires a higher degree of fine tuning. This subtle conspiracy between initial masses and radiative corrections makes this scenario implausible from the point of view of naturalness.

\vspace{0.5cm}
$b)$ $m_{3} \simeq -m_{4}$
\vspace{0.5cm}

In this case, the neutrino mass eigenvalues at $M_Z$ are given by eq.(\ref {eigensame}), with $m_3(\Lambda)=-m_4(\Lambda)=m_0$, while the solar mass splitting induced by the radiative corrections is given by eq.(\ref {oscilA1}). Now we do not expect abrupt changes in the relevant elements of the mixing matrix for the solar splitting (there is no ambiguity in the diagonalization of the mass matrix). To be more precise, the RGEs for these matrix elements read
\bea
\frac{dV_{s3}}{dt}\simeq -\frac{3h_{t}^2}{16\pi ^2} V_{s3} (1-V_{s3}^2-V_{s4}^2),
\eea
(for $V_{s4}$ the RGE is similar with the interchange $3 \leftrightarrow 4$). The BBN bound (\ref{BBNA}) implies that these elements can be considered constant in first approximation, hence $I_{si}\simeq I_{t}^{-V_{si}^2}$. It is apparent from (\ref{oscilA1}) that the splitting will be unacceptably large unless $V_{s3}\simeq V_{s4}$.

For definiteness, we will use the simple texture for the mixing matrix given in \cite{MS2}, although our results will not change essentially if we allow for more complicated textures. This texture almost decouples the $\nu_e,\nu_s-\nu_1,\nu_2$ and the $\nu_{\mu},\nu_{\tau}-\nu_3,\nu_4$ sectors (as suggested by BBN bounds) and slightly rotates the $\nu_{1}-\nu_{3}$ plane by a small angle $\sqrt{2}\theta_{LSND}$ to generate a small mixing between the $\nu_e,\nu_s-\nu_{1}, \nu_{2}$ and $\nu_{\mu},\nu_{\tau}-\nu_{3}, \nu_{4}$ sectors, in order to accommodate the LSND signal. We could also have small rotations in the $\nu_{1}-\nu_{4}$, $\nu_{2}-\nu_{3}$ and $\nu_{2}-\nu_{4}$ planes, but, as stressed before, the main results will remain unchanged. The mixing matrix at $M_Z$ is
\bea
\label{texV}
V(M_Z) \simeq \pmatrix{ \cos\theta_{sol} \sin\sqrt{2} \theta_{LSND} & 0 & \cos\theta_{sol} \cos\sqrt{2} \theta_{LSND} & \sin\theta_{sol} \cr
 \cos\theta_{atm} \cos\sqrt{2} \theta_{LSND} & \sin\theta_{atm} & -\cos\theta_{atm} \sin\sqrt{2} \theta_{LSND} & 0\cr
 -\sin\theta_{atm} \cos\sqrt{2} \theta_{LSND} & \cos\theta_{atm} & \sin\theta_{atm} \sin\sqrt{2} \theta_{LSND} & 0\cr
\sin\theta_{sol} \sin\sqrt{2} \theta_{LSND} & 0 & \sin\theta_{sol} \cos\sqrt{2} \theta_{LSND} & \cos\theta_{sol} \cr
}.
\eea 
Using this mixing matrix and eq.(\ref{oscilA1}), it is easy to see that the renormalization group evolution produces a solar mass splitting given by
\bea
\Delta m_{sol}^2\simeq \Delta m_{LSND}^2 [1- I_t^{-2 (\cos^2 \theta_{sol}-\sin^2 \theta_{sol}\cos^2 \sqrt{2} \theta _{LSND})}]
\eea
It is obvious that in order to obtain a sufficiently small splitting we must require $ \cos^2 \theta_{sol} \simeq \sin^2 \theta_{sol} \cos^2 \sqrt{2} \theta_{LSND}$ (typically, to obtain a mass splitting of the solar order the condition $| \cos^2 \theta_{sol} - \sin^2 \theta_{sol} \cos^2 \sqrt{2} \theta_{LSND}|<10^{-4}$ must be fulfilled). Due to the smallness of $\theta_{LSND}$ we infer that the mixing between $\nu_e$ and $\nu_s$ must be nearly maximal, and adjusted with high accuracy. This discredits the small angle MSW solution to the solar neutrino problem (the only acceptable one) from the point of view of naturalness. The only way out would be, as before, an extremely artificial fine tuning between the initial masses and the renormalization group effects.

\subsubsection{Complex Case}

In the introduction it was explained that when $m_4(\Lambda)=m_3(\Lambda)$ the running quickly drives the mixing matrix to a form in which $Re(U_{s3}^*U_{s4})=0$. In this section we will assume that we end up with a phenomenologically viable mixing matrix and then we will investigate the effect of the running on the mass splitting. 

When $Re(U_{s3}^*U_{s4})=0$, the RGEs for $|U_{s3}|^2$ and $|U_{s4}|^2$ can be written as
\bea
\frac{d}{dt}  |U_{s3}|^2\simeq - \frac {3 h_t^2}{8 \pi^2} (1-|U_{s3}|^2 - |U_{s4}|^2) |U_{s3}|^2 + \frac {3 h_t^2}{8 \pi^2} \frac {m_4-m_3}{m_4+m_3} |U_{s3}|^2 |U_{s4}|^2,
\label{RGEcomplex1}
\eea
(for $|U_{s4}|^2$ the RGE is the same with the interchange $3\leftrightarrow 4$). After imposing the BBN constraint (\ref {BBNA}) we can approximate the above equation by
\bea
\frac{d}{dt}  |U_{s3}|^2\simeq -\frac{d}{dt}  |U_{s4}|^2\simeq \frac {3 h_t^2}{8 \pi^2} \frac {m_4-m_3}{m_4+m_3} |U_{s3}|^2 |U_{s4}|^2.
\label{RGEcomplex2}
\eea
As said before, in the four-neutrino scenarios considered here only the small angle MSW solution is allowed \cite{NLA}, then $|U_{s3}|^2$ is small and by the BBN constraint (\ref {BBNA}), $|U_{s4}|^2$ is close to one. Hence, eq.(\ref{RGEcomplex2}) implies that both $|U_{s3}|^2$ and $|U_{s4}|^2$ are approximately constant with the scale, yielding $I_{s3} \simeq I_t^{-|U_{s3}|^2}$ and $I_{s4} \simeq I_t^{-|U_{s4}|^2}$. Under those well founded hypotheses, the solar splitting induced by the running is
\bea
\Delta m_{sol}^2\simeq \Delta m_{LSND}^2 (1-I_t^{2 (|U_{s4}|^2-|U_{s3}|^2)} ) \sim \Delta m_{LSND}^2  (1-I_t^2),
\label{oscilcomplex}
\eea
where we have neglected terms of the order of $|U_{s3}|^2$ and $1-|U_{s4}|^2$. The numerical results are identical to those after eq.(\ref{oscilA2}), again too large by 3-5 orders of magnitude. We conclude that it is impossible to reconcile a phenomenologically acceptable MNS matrix and an experimentally allowed mass splitting without a fine tuning between initial conditions and renormalization group effects. This permits us to completely discredit the scheme A for the four neutrino spectrum from a theoretical point of view.

\subsection {Scheme B}

Things work quite differently in this case. For the SM, the induced splitting between the heaviest pair, to be identified with the atmospheric oscillation, is given by
\bea
\Delta m_{atm}^2\simeq m_0^2 ~ I_t^2 I_{s4}^2(1 - I_{s3}^2 I_{s4}^{-2}), 
\label{atmosB1}
\eea
with $m_0^2  I_t ^2 I_{s4}^2 =m_4^2(M_Z)  \simeq \Delta m_{LSND}^2$. The BBN constraint (\ref {BBNB}) requires both $|U_{s3}|^2$ and $|U_{s4}|^2$ to be smaller than $10^{-4}$.  This condition is approximately maintained along the running, as can be checked from 
\bea
\frac{d}{dt}\left( |U_{s3}|^2+|U_{s4}|^2\right) \simeq -\frac{3h_{t}^2}{8\pi ^2}(1-|U_{s3}|^2-|U_{s4}|^2) (|U_{s3}|^2+|U_{s4}|^2).
\eea
This allows the following  expansion of eq.(\ref {atmosB1})
\bea
\Delta m_{atm}^2\simeq 2 \Delta m_{LSND}^2 (|U_{s4}|^2 - |U_{s3}|^2) \log I_t.
\label{atmosB2}
\eea
Then, the atmospheric splitting induced by the RGEs is $\simlt 2 \times 10^{-4} \Delta m_{LSND}^2$, too low to account for the atmospheric signal. Inversely, a mass difference of the order of the atmospheric oscillation generated at a high scale $\Lambda$ will not be upset by the radiative corrections.

\begin{figure}
\centerline{\vbox{
\psfig{figure=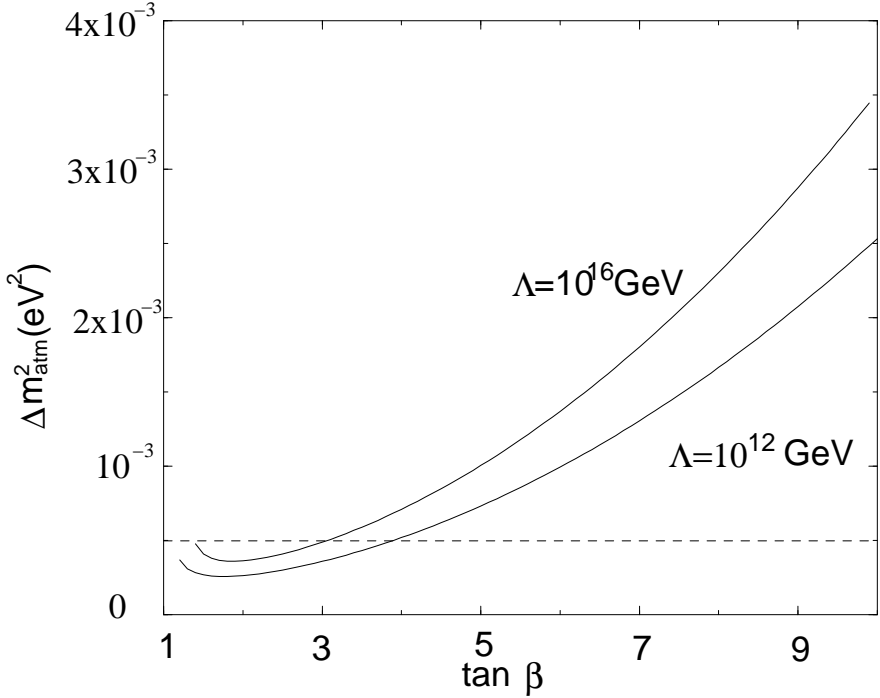,height=10cm,width=10cm,bbllx=3.cm,%
bblly=1.cm,bburx=18.cm,bbury=16.cm}}}
\caption
{\footnotesize
Atmospheric splitting for $\Lambda = 10^{12}$ GeV and $\Lambda =10^{16}$ GeV. The dashed line corresponds to the lower bound for $\Delta m^{2}_{atm}$. }
\end{figure}

For the MSSM case, due to the smallness of $|U_{s3}|^2$ and $|U_{s4}|^2$, the effects induced by the top Yukawa coupling will also be small. However, the effects of the tau Yukawa coupling can be dominant in the large $\tan\beta$ regime \cite{ellis,cein2}, so it is worth considering them in more detail. After including this coupling, the renormalization group equations for the mass eigenvalues are
\be
\frac{d m_i}{dt} \simeq  \left[\frac {3 h_{t}^2}{8 \pi^2 \sin^2 \beta} (1 - |U_{si}|^2)+\frac {h_{\tau}^{2}}{8 \pi^2 \cos^2 \beta}|U_{\tau i}|^2 \right] \;m_{i} ,
\ee
where $i=1,2,3,4$ and $h_{\tau}$ is the tau Yukawa coupling. The induced splitting, in the leading-log approximation, is given by
\be
\Delta m_{atm}^2 \simeq 2\Delta m_{LSND}^2 \left[ k_t ( |U_{s4}|^2
-|U_{s3}|^2) +k_{\tau} (|U_{\tau 3}|^2
-|U_{\tau 4}|^2) \right] \log(\Lambda /M_Z),
\ee
where 
\be
k_t = \frac{3h_t^2}{8 \pi^2 \sin^2 \beta}, ~~~~
k_{\tau} = \frac{h_{\tau}^2}{8\pi^2 \cos^2 \beta}.
\ee
(The leading-log approximation is well justified because $|U_{s3}|^2$, $|U_{s4}|^2$ and $h_{\tau}$ are small.) In fig.(2) it is plotted the maximum splitting reachable by the RG evolution for $\Lambda=10^{12}$ GeV and $\Lambda= 10^{16}$ GeV. We have assumed the most favorable case, i.e., $\Delta m_{LSND}^2=1\; \rm{eV^2}$, $  (|U_{s4}|^2 - |U_{s3}|^2) \sim 10^{-4}$ and $\sin^2 \theta _{atm} \simeq 0.82$, so that $( |U_{\tau 3}|^2-|U_{\tau 4}|^2) \sim 0.42$. From fig.(2) it is apparent that in order to generate the atmospheric splitting radiatively, a moderately large $\tan \beta$ is needed, or a sizeable value of the cut-off.

\section{Conclusions}

We have shown that radiative corrections in models with sterile neutrinos are potentially dangerous in many scenarios. Due to the different interactions of the sterile neutrino, the renormalization group behavior of the mass matrix for the active and sterile neutrinos presents a strong non-universality. This leads to induced splittings proportional to the top Yukawa coupling, in contrast with the three neutrino scenario, where these splittings were found to be proportional to the tau Yukawa coupling. This strong scale dependence of the mass matrix texture has allowed us to make some statements concerning the viability of some neutrino spectra. More precisely, we have analyzed the two schemes favored by the oscillation data, both consisting in two pairs of nearly degenerate neutrinos separated by the LSND gap. Requiring compatibility with the measurements of the abundances of primordial elements produced in Big Bang Nucleosynthesis, we have found that when the heaviest pair corresponds to the solar neutrinos (mainly an admixture of $\nu_{e}-\nu_{s}$) the splitting is not stable under radiative corrections, and the induced one is 3-5 orders of magnitude larger than the observed splitting, making the scenario completely implausible from a theoretical point of view, the only way out being an extreme fine tuning in the high energy parameters. Incidentally, this scenario can be tested in the near future with neutrinoless double-$\beta$ decay experiments \cite {betadecay}.

On the other hand, when the heaviest pair corresponds to the atmospheric neutrinos (mainly $\nu_{\mu}-\nu_{\tau}$), the scenario is safe from radiative corrections due to the small mixing of these mass eigenstates with the sterile neutrino. For the SM, the induced mass differences are lower than the atmospheric splitting (although close to the lower bound for this oscillation) while for the MSSM, there are regions of the parameter space where, beginning with degenerate neutrinos, the atmospheric splitting can be generated radiatively for a moderately large $\tan \beta$ or a sizeable cut-off.

\section*{Acknowledgements}

We would like to thank J. A. Casas and J. R. Espinosa for invaluable discussions and encouragement. The work of A. I. has been supported in part by a Comunidad de Madrid grant.


\end{document}